\documentclass[twocolumn,aps,amsmath,amssymb,prb,superscriptaddress]{revtex4}

\usepackage{graphicx}
\usepackage{amsmath}
\usepackage{bm}

\begin{document}

\title{Spin-orbit interaction and spin relaxation in a two-dimensional electron gas}

\author{M. Studer}
\affiliation{IBM Research, Zurich Research Laboratory,
S\"aumerstrasse 4, 8803 R\"uschlikon, Switzerland}
\affiliation{Solid State Physics Laboratory, ETH Zurich, 8093
Zurich, Switzerland}
\author{S. Sch\"on}
\affiliation{FIRST Center for Micro- and Nanosciences, ETH Zurich,
8093 Zurich, Switzerland}
\author{K. Ensslin}
\affiliation{Solid State Physics Laboratory, ETH Zurich, 8093
Zurich, Switzerland}
\author{G. Salis}
\affiliation{IBM Research, Zurich Research Laboratory,
S\"aumerstrasse 4, 8803 R\"uschlikon, Switzerland}

\date{\today}

\begin{abstract}
Using time-resolved Faraday rotation, the drift-induced spin-orbit
field of a two-dimensional electron gas in an InGaAs quantum well is
measured. Including measurements of the electron mobility, the
Dresselhaus and Rashba coefficients are determined as a function of
temperature between 10 and 80\,K. By comparing the relative size of
these terms with a measured in-plane anisotropy of the
spin-dephasing rate, the D'yakonv-Perel' contribution to spin
dephasing is estimated. The measured dephasing rate is significantly
larger than this, which can only partially be explained by an
inhomogeneous $g$ factor.
 \end{abstract}

\maketitle

The possibility to manipulate spins in semiconductors is a requirement for future spin-based information processing.
\cite{Awschalom2007} Using the spin-orbit (SO) interaction
\cite{Dresselhaus1955,Bychkov1984} is a promising way to
precisely control spin polarization because of its simple principle based on
external gate electrodes. \cite{Datta1990,Hall2006} Manipulation of spins using the SO interaction has been shown in various semiconductor
systems, such as bulk semiconductors~\cite{Kato2004}, two-dimensional
electron gases~\cite{Meier2007} (2DEGs), and even quantum dots
containing only one single electron.~\cite{Nowack2007} On the other
hand, the SO interaction is a source for spin dephasing. In 2DEGs, the
SO interaction induces a linear $k$-dependent splitting.
\cite{Silsbee2004} This splitting gives rise to an effective
magnetic field, leading to dephasing of the polarized electron
spins.~\cite{Dyakonov1972} This effect is known as the D'yakonv-Perel' (DP)
mechanism, and its control through manipulation of the SO interaction has been
proposed~\cite{Schliemann2003} as an alternative to the ballistic
spin transistor.~\cite{Datta1990} A careful engineering of the SO
interaction is therefore crucial for using it to manipulate the spin.

In a 2DEG at intermediate temperatures, it is often assumed that the
spin decay is governed by the DP
mechanism.~\cite{Zutic2004,Malinowski2000} Based on this assumption,
information on the SO interaction in semiconductor quantum wells
(QWs) was obtained from measurements of the spin-dephasing
rate.~\cite{Averkiev2006,Eldridge2007,Larionov2008} An independent
measurement of the relative size of the SO interaction in
(110)-grown QWs using the photogalvanic effect
 has been described in Ref. \onlinecite{Belkov2008} and compared to the spin decay time.
 In this paper, we report on quantitative and independent measurements of the SO interaction and the spin-dephasing rate in an InGaAs QW, utilizing time-resolved
Faraday rotation.  In a further development of the method described
in Ref.~\onlinecite{Meier2007}, a well-defined current is applied in
the 2DEG using Ohmic contacts and a mesa structure (in
Ref.~\onlinecite{Meier2007}, the electron drift was induced by an ac
voltage applied to Schottky contacts in an unstructured 2DEG). The
drifting electrons see an effective SO magnetic field, in the
following referred to as drift SO field. The sizes of its two
contributions, the Rashba~\cite{Bychkov1984} and the
Dresselhaus~\cite{Dresselhaus1955} fields, are determined as a
function of temperature $T$ from the measured influence of the
in-plane electron drift velocity on the spin precession. Comparing
our results with measured spin-dephasing rates and their in-plane
anisotropy, we find that DP is not the only mechanism for spin decay
in our samples at $T$ between 10 and 80\,K.

\begin{figure}
\includegraphics[width=80mm]{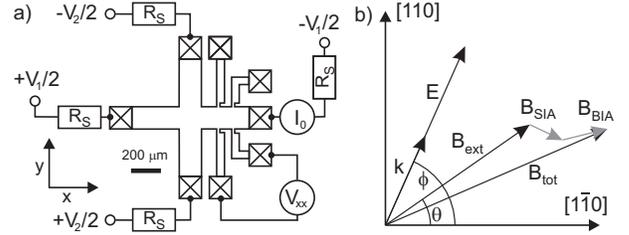}
\caption{\label{fig:fig1} (a) The InGaAs/GaAs QW is shaped into a cross with $150$-$\mu$m-wide arms contacted by Ohmic contacts. Additional
contacts on one arm of the cross allow a four-point measurement of the voltage drop $V_{xx}$ and
the determination of electron sheet density and mobility. (b) $\theta$
and $\phi$ are the angles of the magnetic and electric fields with
respect to the [$1\overline{1}0$] axis. $\boldsymbol{B}_\textrm{tot}$
is the vector sum of the external magnetic field and the two SO
effective magnetic field contributions.}
\end{figure}

The 2DEG we use in this work is located in an
In$_{0.1}$Ga$_{0.9}$As/GaAs QW. Electrons are confined to a
20-nm-thick In$_{0.1}$Ga$_{0.9}$As layer that is $n$-doped
(3$\times10^{16}$ cm$^{-3}$) to ensure a small electron scattering
time such that we are in the dirty limit of the SO
interaction\cite{Silsbee2004}, where the frequency
${\omega_\textrm{SO}}$ of spin precession about the SO fields is
small compared with the momentum scattering rate $1/\tau_{p}$
($\omega_\textrm{SO}\tau_{p}\approx 10^{-5}$ for our samples). On
both sides of this layer, there is a 10-nm-thick GaAs spacer layer
and a 10-nm-thick layer of $n$-doped GaAs. A 10\,nm cap of undoped
GaAs completes the structure, grown by molecular beam epitaxy and
forming a 2DEG 40\,nm below the surface. We use wet etching to
pattern a cross-shaped mesa as shown in Fig. \ref{fig:fig1}(a) and
create standard AuGe Ohmic contacts in the four ends of the cross.
Four additional contacts on one arm of the cross allow its use as a
Hall bar to determine the resistivity and carrier density of the
2DEG. Two samples with the same structure are glued into one chip
carrier, whereby one sample is rotated by 90$^\circ$ to allow the SO
interaction to be measured in one cool-down. At 40\,K, the two-point
resistance of the crosses in $\boldsymbol{x}$ or $\boldsymbol{y}$
direction is 4.1\,k$\Omega$. We use additional resistors
$R_{s}=4.7\,$k$\Omega$ to compensate for small variations in the
contact resistance and apply voltages $V_{1}=V_A\cos(\phi)$ and
$V_{2}=V_{A}\sin(\phi)$ as shown in Fig.~\ref{fig:fig1}(a).  All
angles are given with respect to the
$x$ axis along $[1\overline{1}0]$, as seen in
Fig.~\ref{fig:fig1}(b). We obtain the resistivity of the 2DEG during
optical experiments monitoring the ratio of the voltage drop
$V_{xx}$ and the current $I_\textrm{0}$ through one arm of the mesa
[see Fig.~\ref{fig:fig1}(a)], and measure a value of 770 $\Omega$/sq
at 40\,K. The voltages $V_1$ and $V_2$ create an electric field
$\boldsymbol{E}$ in the center of the cross in direction $\phi$ and
with an amplitude proportional to $V_{A}$. Because $V_{xx}$ is
monitored, the corresponding component of $\boldsymbol{E}$ can be
determined directly (see below). The electric field shifts the Fermi
circle by an amount of $\delta \boldsymbol{k}=m^* \mu \boldsymbol{E}
/ \hbar$, where $m^*$ is the effective electron mass, $\mu$ is the
electron mobility, and $\hbar$ is Planck's constant divided by
$2\pi$. In the dirty SO limit, the shift induces drift SO fields
that can be expressed as \cite{Silsbee2004}
\begin{equation}\label{eq:SIABIA}
   \boldsymbol{B}_{\textrm{SIA}} = \frac{2\alpha}{g \mu_B} \binom{\quad \delta k_y}{-\delta k_x}, \quad
   \quad
   \boldsymbol{B}_{\textrm{BIA}} = \frac{2\beta}{g \mu_B} \binom{\delta k_y}{\delta k_x},
\end{equation}
with $\delta\boldsymbol{k} = (\delta k_x, \delta k_y)$, $g$ is the
electron $g$ factor, $\mu_B$ is the Bohr magneton, and $\alpha$ and
$\beta$ are the Rashba and Dresselhaus spin-orbit coefficients,
respectively. The Rashba field has its origin in the structure
inversion asymmetry (SIA) due to nonuniform doping on both sides of
the QW and the presence of the surface on one side of the QW. The
Dresselhaus field is a consequence of the bulk inversion asymmetry
(BIA) of the zinc-blende structure.
Cubic Dresselhaus terms do not change the linearity of
$\boldsymbol{B}_{\textrm{BIA}}$ in $\delta k$, but introduce a
correction of $1-\frac{1}{4}k_{F}^2/\langle k_z^2\rangle$, where
$k_\textrm{F}$ is the Fermi wave number and $\langle k_z^2\rangle$
the expectation value of the squared wave number along the growth
direction $z$. Taking a sheet density of $5.2 \times 10^{15}$
m$^{-2}$ (see below) and approximating $\langle k_z^2\rangle$ by
$(\pi/w)^2$, where $w=20$\,nm is the QW width, we obtain
$k_{F}^2\approx 1.32 \langle k_z^2\rangle$. This gives a correction
in $\beta$ of about than 35\%, which will be neglected in the
following.

An external magnetic field ${B}_\textrm{ext}$ is applied in the
direction $\theta$ as seen in Fig.~\ref{fig:fig1}(b). The angles
$\theta$ are $90^\circ$ for sample\,1 and $180^\circ$ for sample\,2.
A transverse electron polarization precesses coherently about the
vector sum\cite{Kalevich1990, Engel2007, Duckheim2007} of
$\boldsymbol{B}_\textrm{ext}$ and the drift SO fields defined in
Eq.~(\ref{eq:SIABIA}) with a frequency given by the modulus of this
total field vector. If $B_\textrm{ext} \gg B_{\textrm{SIA}},
B_{\textrm{BIA}}$, the total field can be approximated as
\cite{Meier2007}

\begin{equation}\label{eq:AB}
\begin{split}
B_\textrm{tot}(\theta,\phi)  \approx & B_\textrm{ext}+(B_\textrm{BIA}+ B_\textrm{SIA})\cos{\theta}\sin{\phi} + \\
&(B_\textrm{BIA}-B_\textrm{SIA})\sin{\theta}\cos{\phi}.\\
\end{split}
\end{equation}

Because of the different angular dependencies of the Rashba and
Dresselhaus SO fields, the two contributions can be distinguished.
We use time-resolved Faraday rotation to determine the Larmor
frequency $\Omega_\textrm{L}=g\mu_{B}B_\textrm{tot}/\hbar$ of the
spins precessing about $\boldsymbol{B}_\textrm{tot}$. For this, the
output of a pulsed Ti:sapphire laser with a repetition rate of
80\,MHz is split into a pump and a probe beam. The pump/probe
intensity ratio is 10/1 and, unless stated otherwise, the pump power
is 500\,$\mu$W, focused onto a spot 30\,$\mu$m in diameter.  The
circularly polarized pump pulse is tuned to the absorption edge of
the QW at 1.44\,eV and creates a spin polarization along the growth
axis of the QW. With a pump power of 500\,$\mu$W and assuming an
absorption of about 1\%\cite{Miller1982} we obtain a photo-excited
carrier concentration on the order of a few $10^{10}$\,cm$^{-2}$,
which is more than a magnitude smaller than the equilibrium carrier
sheet density in the QW (see below). The Faraday rotation of the
linear polarization axis of the probe pulse transmitted is
proportional to the spin polarization along the QW growth axis.
Changing the delay $\Delta t$ between pump and probe reveals the
spin dynamics of the system, and the Faraday rotation angle can be
described by $A \exp(-\Delta t / T_2^*) \cos(\Omega_{L} \Delta t)$.
Here, $A$ is the amplitude of the Faraday signal and $T_2^*$ the
spin dephasing time. A measurement of $\Omega_{L}$ in a known
magnetic field reveals an electron $g$ factor of $-0.29$, assuming
that the $g$ factor is negative.

\begin{figure}
\includegraphics[width=80mm]{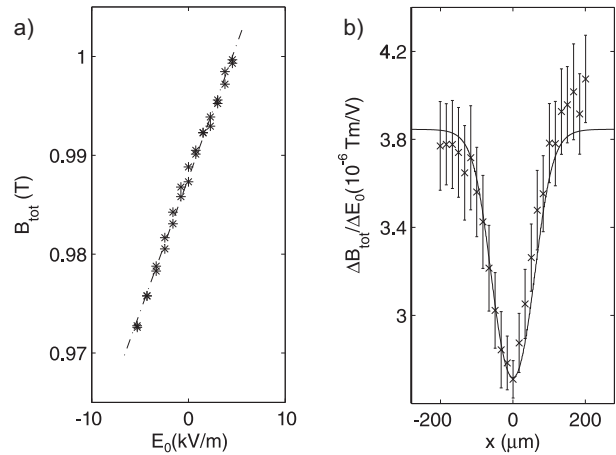}
\caption{\label{fig:fig2} (a) $|\boldsymbol{B}_\textrm{tot}|$ in the
center of the cross (symbols) as a function of the measured electric field.
The slope of the fitted line is  2.71$\times 10^{-6}$\,Tm/V. (b)
Dependence of the fitted slope on the position $x$ (crosses). Assuming a constant $\alpha-\beta$ of $4.9\times 10^{-14}$\,eVm, a simulation of the electric field reproduces the measured values (solid line).}
\end{figure}

Figure~\ref{fig:fig2}(a) shows $B_\textrm{tot}$ measured in the
center of the cross as a function of the electric field
$E_0=V_{xx}/l$ between the two contacts in the right arm. These
contacts are separated by a distance of $l$\,=\,100\,$\mu$m. $V_{2}$
is set to ground. The temperature of the sample is 40\,K, and
$B_\textrm{ext}$ is oriented along the [110] direction, therefore
$\theta = \phi = 90^\circ$. The data in Fig.~\ref{fig:fig2}(a)
contain values from sweeps of $V_\textrm{1}$ up and down. The up and
down sweeps fit nicely to a straight line, showing that we can
exclude a drift of $\Omega_{L}$ over time, which might be caused by
nuclear polarization or drift of laser power or temperature.

As a result of the sample geometry and the grounded contacts to the
left and the right of the center, the electric field in the center
of the cross is reduced as compared to the value $E_0$ measured in
the arm. By scanning the laser beam along the $x$-axis and centered
in the $y$-direction [see Fig.~\ref{fig:fig1}(a)], we obtain a cross
section of the drift SO field that is related to the electric field
distribution.\cite{Meier2008} The resulting slopes of the linear
fits of $B_\textrm{tot}$ vs. $E_0$ are shown as a function of $x$ in
Fig.~\ref{fig:fig2}(b). We see a pronounced dip in the center of the
cross, which is explained by the reduced electric field. The solid
line in Fig.~\ref{fig:fig2}(b) represents the solution of a numeric
simulation of the electrostatics using a two-dimensional partial
differential equation solver. Assuming that $\alpha$ and $\beta$ are
independent of the position and using the measured mobility (see
below), the only fit parameter left is the difference $\alpha -
\beta$. The measurement and the simulation are consistent and show
that the electric field $E$ in the middle of the cross is 0.71 times
smaller than the measured value $E_0$. This correction is taken into
account in the following when indicating electric-field values.

\begin{figure}
\includegraphics[width=80mm]{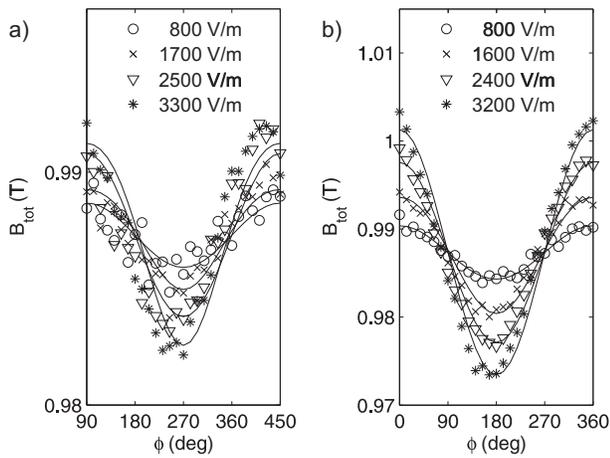}
\caption{\label{fig:fig3}  $B_\textrm{tot}$ as a function of the
direction $\phi$ of the in-plane electric field for different amplitudes $E$ and for two external magnetic field
directions (a) $\theta = 180^\circ$ and (b) $\theta = 90^\circ$. The solid lines are fits using Eq.\,(\ref{eq:AB}). The
data are obtained at $T=20$\,K.}
\end{figure}

To disentangle $\alpha$ and $\beta$, we position the laser spot in
the center of the cross and rotate the electric field. We did such
experiments for different amplitudes of the electric field up to
3.3\,kV/m and for two different configurations of
$\boldsymbol{B}_\textrm{ext}$. In Fig.~\ref{fig:fig3}(a), $\theta$ =
$180^\circ$, and in Fig.~\ref{fig:fig3}(b), $\theta$ = $90^\circ$.
The data are obtained at 20\,K. $B_\textrm{tot}$ oscillates in
$\phi$ with an amplitude that is proportional to $\alpha+\beta$ for
$\theta=180^\circ$ and to $\alpha-\beta$ for $\theta=90^\circ$. The
difference in the amplitude for the two cases (note the different
scales) shows that $B_\textrm{SIA}$ and $B_\textrm{BIA}$ are
comparable in relative strength and that the interplay of the two SO
effects gives rise to an anisotropic spin splitting in $k$ space.
The solid lines are a fit to the data using Eq.~(\ref{eq:AB}). Small
deviations of the data from theory in the $\phi$ direction could
result from a slight accidental off-center position of the laser
spot.

To calculate the SO coefficients $\alpha$ and $\beta$ from the
measured $B_\textrm{SIA}$ and $B_\textrm{BIA}$, we need a value for
the drift momentum of the electrons. This is obtained from a Hall
measurement of the sheet resistivity $\rho$ and the sheet carrier
density $n$. We calculate the mobility $\mu$ using $\mu=1/ne\rho$,
$e$ being the electron charge. In the dark, the resistivity of the
2DEG is approximately 1000 $\Omega$/sq and decreases to 770
$\Omega$/sq under conditions of the optical measurements. The Hall
sheet densities are $5.8\times 10^{15}$ m$^{-2}$ under illumination
and $4.5\times 10^{15}$ m$^{-2}$ in the dark. These densities are
constant from 10 to 80\,K. From Shubnikov--de Haas oscillations at
$T<20$\,K, we get a sheet density of $5.2 \times 10^{15}$ m$^{-2}$
under optical illumination. A small parallel conductivity of a
doping layer could explain the difference in the two numbers. Such a
parallel conductivity does not influence the optical experiments as
these electrons do not contribute to the Faraday signal.

\begin{figure}
\includegraphics[width=80mm]{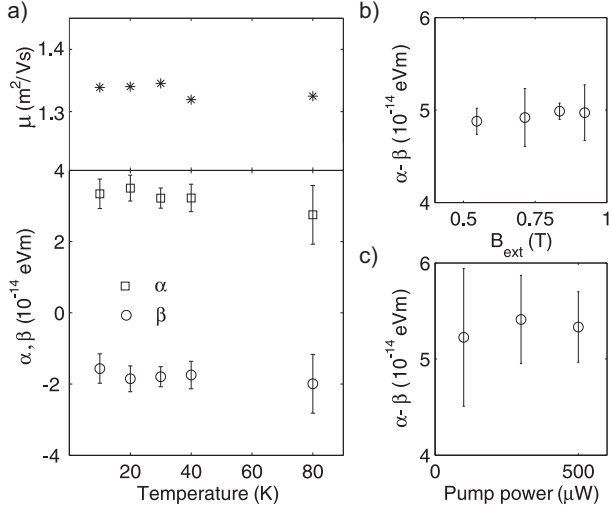}
\caption{\label{fig:fig4} (a) Measured $\mu$, $\alpha$ and
$\beta$ vs. $T$. (b) Measured $\alpha-\beta$ shows no significant
dependence on $B_\textrm{ext}$.  (c) Pump-power dependence of
$\alpha-\beta$. The data in (a) and (b) were obtained with a pump
power of 500\,$\mu$W, the data in (b) and (c) at $T$\,=\,40 K.}
\end{figure}

The mobilities extracted from the Hall measurement under
illumination are shown in the upper inset of Fig.~\ref{fig:fig4}(a).
The mobility does not change significantly over the temperature
range measured. Using the results of the transport measurements and
assuming that $m^*=0.058$, we can calculate $\delta k$ and use
Eq.~(\ref{eq:SIABIA}) to obtain $\alpha$ and $\beta$ for all the
temperatures measured. The results are displayed in
Fig.~\ref{fig:fig4}(a). Error bars show a $2\sigma$ confidence
interval. The wafer used for this work is the same as the one used
for sample\,3 in Ref.~\onlinecite{Meier2007}. We measure values for
$\alpha$ and $\beta$ that are by a factor of 2-3 smaller, which we
attribute to a more precise determination of the electric field in
this work. In addition, different wafer processing and oxidation of
the wafer surface over time might influence the SO coefficients
measured in this shallow 2DEG. Variations in $\alpha$ and $\beta$
for subsequent cool downs are within the error bar.
Note that we extrapolate the mobility and the electric field in the center of the cross from a transport measurement away from the center. We cannot exclude that we underestimate the absolute values for $\alpha$ and $\beta$ because of a reduced electron drift momentum that might result from, e.g., screening by the optically excited charge carriers.
The model of Vi\~na \cite{Vina1984} with the results and the
parameters used by H\"ubner et al. \cite{Huebner2006} predicts the
$T$ dependence of the band parameters, from which we estimate the
$T$ dependence of $\alpha$ and $\beta$ using
$\boldsymbol{k}\cdot\boldsymbol{p}$ theory.\cite{Winkler2003} The
calculated $T$-induced change in $\alpha$ and $\beta$ between 10 and
80\,K is in the sub-percentage range and thus much smaller than our
measurement error. Interestingly, in
Ref.~[\onlinecite{Eldridge2007}] a linear increase in $\alpha$ with
$T$ for higher $T$ was observed on a [110] QW, which the authors
could not explain with $\boldsymbol{k}\cdot\boldsymbol{p}$ theory.

We find no dependence of the SO coefficients on $B_\textrm{ext}$, in
agreement with our assumption that the precession frequency is given
by the modulus of the vector sum of $B_\textrm{ext}$ and the drift
SO fields given in Eq.~(\ref{eq:SIABIA}). Figure~\ref{fig:fig4}(b)
shows a measurement of $\alpha-\beta$ vs $B_\textrm{ext}$. The
insensitivity of the result on $B_\textrm{ext}$ excludes a
significant admixture of a $k$ dependent and anisotropic $g$ factor,
as was stipulated in Ref.~\onlinecite{Margulis1983}. To test the
reliability of our method, we also checked whether a lower pump
power will influence the outcome of the measurement. This could
occur from, e.g., a population of higher energy states with larger
pump power. We found, however, that $\alpha-\beta$ does not depend
significantly
on the pump power, as seen in Fig.~\ref{fig:fig4}(c).\\

\begin{figure}
\includegraphics[width=80mm]{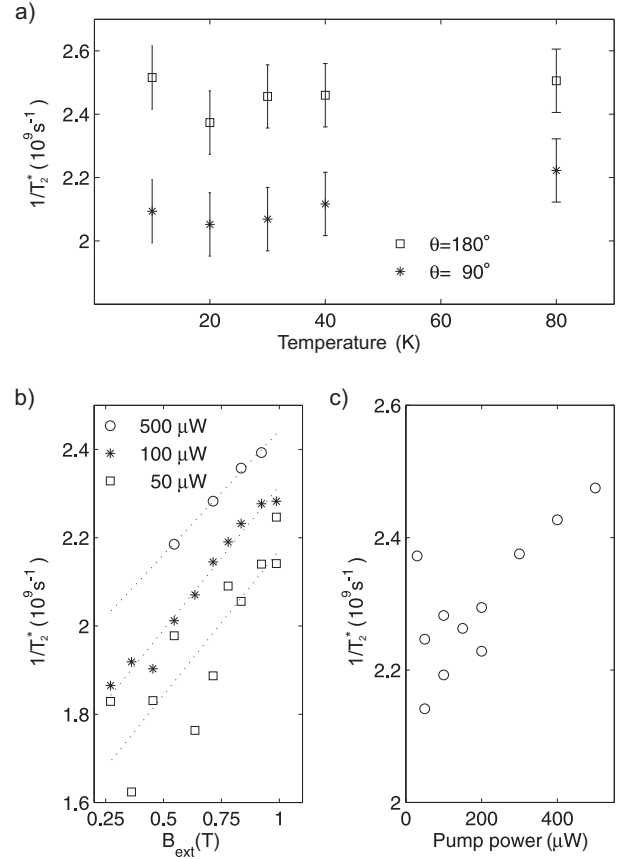}
\caption{\label{fig:fig5}(a) Spin dephasing rate $1/T_2^*$ for two
different in-plane directions $\theta$ vs. $T$ from 10 to 80 K. The average of three measurements with 50\,$\mu$W pump power was used.
 (b) $1/T_2^*$ vs. $B_\textrm{ext}$ and linear fits for three different pump powers.
 The data is measured in the  $\theta = 90^\circ$ configuration
 at 10\,K, and at 40\,K for the 500 $\mu$W case. (c) Pump-power dependence of $1/T_2^*$
measured at 10\,K in the $\theta = 90^\circ$ configuration.}
\end{figure}

When $\alpha$ and $\beta$ are of similar magnitude, the spin lifetime
is strongly anisotropic with respect to the direction of $\boldsymbol{B}_\textrm{ext}$
 in the plane of the 2DEG.~\cite{Averkiev1999,Averkiev2006,Schliemann2003,Duckheim2007,StichC2007,Larionov2008}
 This anisotropy is a consequence of the DP mechanism because the spins precess about
 a SO field whose direction becomes independent of $k$ for $\alpha\approx\beta$.
 From the measured anisotropy in $T_2^*$ and the relative size of $\alpha$ and $\beta$,
 we estimate the contribution from the DP mechanism to the spin dephasing. In Fig.~\ref{fig:fig5}(a),
 the spin relaxation rate 1/$T_2^*$ is plotted as a
function of $T$ for the two orientations of
$B_\textrm{ext}$=0.99\,T. The $T$ dependence is rather small, and we
can clearly see an anisotropy of 1/$T_2^*$, confirming the
anisotropic spin splitting in our system. Because our 2DEG is well
in the dirty SO limit, we can use the motional narrowing limit of
the DP mechanism,\cite{Silsbee2004} where the spin dephasing due to
$|\boldsymbol{k}|$-dependent SO fields is decreased by
spin-preserving scattering. This gives the following expressions for
the anisotropic spin decay rates:
\begin{equation}
\label{eq:tau} \frac{1}{\tau_{z}}=C(\alpha^2+\beta^2), \qquad
\frac{1}{\tau_{x,y}}=\frac{C}{2}(\alpha\pm\beta)^2. \qquad
\end{equation}
Here, $\tau_{x,y,z}$ are the relaxation times of spins oriented
along $x$, $y$, or $z\|[001]$. $C$ is a constant that depends on
$T$, Fermi energy, scattering time, and the scattering
mechanism.\cite{Kainz2004} If we apply a large external magnetic
field ($\Omega_{L} \tau_{x,y,z}\gg1$ ), we can write the DP
spin-dephasing rate as \cite{Larionov2008}
\begin{equation}
\label{eq:tauandB}
\frac{1}{\tau_\textrm{DP}(\theta)}=\frac{1}{2}(\frac{1}{\tau_z}+
\frac{\sin^2\theta}{\tau_x}+\frac{\cos^2\theta}{\tau_y}).
\end{equation}
For the difference, we get
$\frac{1}{\tau(90)}-\frac{1}{\tau(180)}=C\alpha\beta$ and read a
value of about $0.4 \times 10^9$ s$^{-1}$ in Fig.~\ref{fig:fig5}(a)
for the difference. Using the measured values for $\alpha$ and
$\beta$, we get $C = 6.6\times 10^{35}$\,m$^{-2}$eV$^{-2}$s$^{-1}$.
From Eqs.~(\ref{eq:tau}) and (\ref{eq:tauandB}), this yields
relaxation rates for DP of about $0.6\times 10^9$ s$^{-1}$ for
$\theta=180^\circ$, and of $1.0\times 10^9$ s$^{-1}$ for
$\theta=90^\circ$.
In Eq.~(\ref{eq:tau}), it is assumed that the SO splitting is linear
in $k$. As mentioned earlier, we are in a regime where
$k_{F}^2\approx\langle k_z^2\rangle$. Taking into account the cubic
Dresselhaus terms~\cite{Kainz2004} we find only a small correction
to the values for the spin relaxation rate obtained above.
As the total 1/$T_2^*$ lies between 2.1$\times 10^9$ and 2.5$\times
10^9$ s$^{-1}$, other spin-dephasing mechanism must be present in
our sample.

To exclude optical recombination as a source of decay of the
Faraday signal, we measured the time-resolved
reflection,\cite{Malinowski2000} which exponentially decays with a
decay time of less than 100\,ps (data not shown). Interpreting this time
as the electron-hole recombination time provides evidence that the
spin polarization, which is observable over a much longer time
scale in the Faraday signal, must get imprinted onto the equilibrium
electrons in the QW conductance band through recombination of
unpolarized electrons and holes.\cite{Kikkawa1998} It is therefore
justified to interpret the decay time of the Faraday signal fitted
in a window from 80 to 1000\,ps as the decay time $T_2^*$.

From the dependence of $T_2^*$ on $B_\textrm{ext}$, information on
the mechanism of spin dephasing can be obtained. The DP spin
dephasing rate does not depend on $B_\textrm{ext}$ in the motional
narrowing regime and for $\Omega_{L} \tau_p \ll
1$.\cite{Ivchenko1973} In contrast to this, a $B$-dependence that is
intrinsic to the DP mechanism is observed in high-mobility
samples.\cite{StichC2007} In our low-mobility samples, we find a
linear increase in $T_2^*$ with $B_\textrm{ext}$, as shown in
Fig.~\ref{fig:fig5}(b). Such a linear $B$-dependence is evidence of
an inhomogeneous dephasing due to a variation $\Delta g$ of the $g$
factor in the area of the 2DEG probed, described by a dephasing rate
$1/\tau_{{\Delta g}}$,\cite{Kikkawa1998,Chen2007}
\begin{equation}
\label{deltag} \frac{1}{\tau_{{\Delta g}}} = \frac{\Delta g
\mu_{{B}}B_\textrm{ext}}{2\hbar}.
\end{equation}
We suspect that the $g$ factor variation could be a consequence of
the in-well doping. That the sample is rather inhomogeneous is also
seen in a photoluminescence experiment, in which we observe a broad
luminescence peak from the QW (not shown) with a full width at half
maximum of about 20\,meV. We used different pump intensities and
find a similar slope for the dashed linear fits in
Fig.~\ref{fig:fig5}(b). The $500\,\mu$W measurement was done at
40\,K, and the other two at 10\,K.
 From these data, we conclude that
$\Delta g$ is quite constant for different pump powers. From the
slopes in Fig.~\ref{fig:fig5}(a) and using Eq.~(\ref{deltag}), we
obtain $\Delta g=0.014$. Unexpectedly in a doped sample with fast
electron-hole recombination, the overall spin-relaxation rate
increases with increasing pump power; see Fig.~\ref{fig:fig5}(c). In
a high-mobility sample, a decrease in the spin relaxation-rate with
increasing initial spin polarization has been observed, which goes
into the opposite direction.\cite{StichB2007} In an attempt to
minimize this pump-power dependence, we used a low pump power of 50
$\mu$W for the measurement in Fig.~\ref{fig:fig5}(a).

\begin{table}
\begin{tabular}{cccccc}
\hline
\hline
Relaxation mechanism & DP &  $\Delta g$ & Sum & Measured\\
  \hline
  $\theta=90^\circ$ (1/s)$\times10^9$& 0.6 & 0.7 & 1.3$\pm$0.2 & 2.1$\pm$ 0.1\\
  $\theta=180^\circ$ (1/s)$\times10^9$& 1.0 & 0.7 &  1.7$\pm$0.2 & 2.5$\pm$ 0.1\\
\hline
\hline
\end{tabular}
\caption{\label{table:table1} Contributions to the spin decay rate
1/$T_2^*$ at 30\,K, as estimated from the measured anisotropy of
$1/T_2^*$, SO constants $\alpha$ and $\beta$, the $B$-dependence of
$1/T_2^*$, and the electron mobility. Measured values are obtained at a pump power of 50\,$\mu$W.}
\end{table}

Table~\ref{table:table1} summarizes the contributions to the
anisotropic 1/$T_2^*$ for $T=30$\,K. The calculated sum of the
relaxation rate is by about $0.8\times 10^9$\,s$^{-1}$ lower than
the measured value. This discrepancy indicates the presence of
another spin-dephasing mechanism. A possible candidate is a random
SO field originating from the Coulomb potential of ionized dopants
or from surface roughness of the QW. It has been pointed out that
such spatial fluctuations might limit the spin lifetime for
symmetric (110) QWs (Ref.\,\onlinecite{Karimov2003}) or in the case
where $\alpha=\beta$.\cite{Sherman2005} The importance of this
effect is probably smaller in our samples where both $\alpha$ and
$\beta$ are finite but not equal in size. In a small-gap
semiconductor, the Elliott-Yafet (EY) mechanism contributes to the
spin dephasing.\cite{Elliott1954} By estimating the importance of
the EY mechanism\cite{Zutic2004} in our sample using the measured
mobility and the known band parameters, we obtain a spin-relaxation
rate on the order of 5$\times10^7$\,s$^{-1}$. This is negligibly
small, but there are indications\cite{Tackeuchi1999} that the EY
spin-dephasing rate might be larger than estimated with the equation
derived for a bulk semiconductor.

The weak variation in $1/T_2^*$ with $T$ shown in
Fig.~\ref{fig:fig5}(a) can be understood as a consequence of little
temperature dependence of the individual contributions to $1/T_2^*$.
As pointed out in Ref.\,\onlinecite{Kainz2004}, the DP dephasing
rate depends only weakly on $T$ in the degenerate regime and in the
intermediate temperature range, apart from its proportionality to
the electron scattering time. As our mobility is quite constant over
the temperature range measured, we do not expect large variations.
We observe no evidence of a dependence of the $g$ factor spread on
$T$. For a degenerate electron density and a constant mobility, also
the $T$ dependence of the EY mechanism should be small. The observed
weak $T$ dependence is therefore not surprising, and has also been
observed in other experiments. \cite{Malinowski2000}

To conclude, we have measured the SO interaction coefficients
$\alpha$ and $\beta$ as a function of $T$ and find no significant
$T$ dependence. From $\alpha$ and $\beta$, the measured Hall
mobility and the anisotropy in $1/T_2^*$, we estimate the
contribution from DP spin dephasing, and find that DP alone cannot
explain the measured $1/T_2^*$. From a linear increase in $1/T_2^*$
with $B_\textrm{ext}$, we identify an inhomogeneous broadening from
a spread in the electron $g$ factor. These effects do not account
for all of the measured spin-dephasing rate. We speculate that EY or
an inhomogeneous SO field might induce an additional isotropic
contribution. A more detailed study of the nature of the elastic and
inelastic electron-scattering mechanisms involved might facilitate
an exact attribution to the different decay mechanisms.

We gratefully acknowledge helpful discussions with T. Ihn and I.
Shorubalko and thank B. K\"ung for evaporating contact metals. This
work was supported by the CTI and the SNSF.

\bibliographystyle{prsty}

\begin{thebibliography}{10}

\bibitem{Awschalom2007}
D.~D. Awschalom and M.~E. Flatt\'{e}, Nat. Phys. {\bf 3},  153
(2007).

\bibitem{Dresselhaus1955}
G. Dresselhaus, Phys. Rev. {\bf 100},  580  (1955).

\bibitem{Bychkov1984}
Y.~A. Bychkov and E.~I. Rashba, J. Phys. C {\bf 17},  6039  (1984).

\bibitem{Datta1990}
S. Datta and B. Das, Appl. Phys. Lett. {\bf 56},  665  (1990).

\bibitem{Hall2006}
K.~C. Hall and M.~E. Flatt\'{e}, Appl. Phys. Lett. {\bf 88},  162503
(2006).

\bibitem{Kato2004}
Y. Kato, R.~C. Myers, A.~C. Gossard, and D.~D. Awschalom, Nature
{\bf 427},  50
   (2004).

\bibitem{Meier2007}
L. Meier, G. Salis, I. Shorubalko, E. Gini, S. Sch\"on, and K.
Ensslin, Nature
  Phys. {\bf 3},  650  (2007).

\bibitem{Nowack2007}
K.~C. Nowack, F.~H.~L. Koppens, Y.~V. Nazarov, and L.~M.~K.
Vandersypen,
  Science {\bf 318},  1430  (2007).

\bibitem{Silsbee2004}
R.~H. Silsbee, J. Phys. Condens. Matter {\bf 16},  R179  (2004).

\bibitem{Dyakonov1972}
M.~I. D'yakonov and V.~I. Perel', Sov. Phys. Solid State {\bf 13},
3023
  (1972).

\bibitem{Schliemann2003}
J. Schliemann, J.~C. Egues, and D. Loss, Phys. Rev. Lett. {\bf 90},
146801
  (2003).

\bibitem{Zutic2004}
I. \ifmmode \check{Z}\else \v{Z}\fi{}uti\ifmmode~\acute{c}\else
\'{c}\fi{}, J.
  Fabian, and S. Das~Sarma, Rev. Mod. Phys. {\bf 76},  323  (2004).

\bibitem{Malinowski2000}
A. Malinowski, R.~S. Britton, T. Grevatt, R.~T. Harley, D.~A.
Ritchie, and
  M.~Y. Simmons, Phys. Rev. B {\bf 62},  13034  (2000).

\bibitem{Averkiev2006}
N.~S. Averkiev, L.~E. Golub, A.~S. Gurevich, V.~P. Evtikhiev, V.~P.
  Kochereshko, A.~V. Platonov, A.~S. Shkolnik, and Y.~P. Efimov, Phys. Rev. B
  {\bf 74},  033305  (2006).

\bibitem{Eldridge2007}
P.~S. Eldridge, W.~J.~H. Leyland, P.~G. Lagoudakis, O.~Z. Karimov,
M. Henini,
  D. Taylor, R.~T. Phillips, and R.~T. Harley, Phys. Rev. B {\bf 77},  125344
  (2008).

\bibitem{Larionov2008}
A.~V. Larionov and L.~E. Golub, Phys. Rev. B {\bf 78},  033302
(2008).

\bibitem{Belkov2008}
V.~V. Bel'kov, P. Olbrich, S.~A. Tarasenko, D. Schuh, W.
Wegscheider, T. Korn,
  C. Sch\"{u}ller, D. Weiss, W. Prettl, and S.~D. Ganichev, Phys. Rev. Lett.
  {\bf 100},  176806  (2008).

\bibitem{Kalevich1990}
V.~K. Kalevich and V.~L. Korenev, JETP Lett. {\bf 52},  230  (1990).

\bibitem{Engel2007}
H.-A. Engel, E.~I. Rashba, and B.~I. Halperin, Phys. Rev. Lett. {\bf
98},
  036602  (2007).

\bibitem{Duckheim2007}
M. Duckheim and D. Loss, Phys. Rev. B {\bf 75},  201305  (2007).

\bibitem{Miller1982}
D.~A.~B. Miller, D.~S. Chemla, D.~J. Eilenberger, P.~W. Smith, A.~C.
Gossard,
  and W.~T. Tsang, Appl. Phys. Lett. {\bf 41},  679  (1982).

\bibitem{Meier2008}
L. Meier, G. Salis, E. Gini, I. Shorubalko, and K. Ensslin, Phys.
Rev. B {\bf
  77},  035305  (2008).

\bibitem{Vina1984}
L. Vi\~na, S. Logothetidis, and M. Cardona, Phys. Rev. B {\bf 30},
1979
  (1984).

\bibitem{Huebner2006}
J. H\"ubner, S. Dohrmann, D. H\"agele, and M. Oestreich,
arXiv:cond-mat/0608534
   (2006).

\bibitem{Winkler2003}
R. Winkler, {\em Spin-Orbit Coupling Effects in Two-Dimensional
Electron and
  Hole Systems} (Springer, Berlin, 2003).

\bibitem{Margulis1983}
A.~D. Margulis and V.~A. Margulis, Sov. Phys. Solid State {\bf 25},
918
  (1983).

\bibitem{Averkiev1999}
N.~S. Averkiev and L.~E. Golub, Phys. Rev. B {\bf 60},  15582
(1999).

\bibitem{StichC2007}
D. Stich, J.~H. Jiang, T. Korn, R. Schulz, D. Schuh, W. Wegscheider,
M.~W. Wu,
  and C. Sch\"{u}ller, Phys. Rev. B {\bf 76},  073309  (2007).

\bibitem{Kainz2004}
J. Kainz, U. R\"ossler, and R. Winkler, Phys. Rev. B {\bf 70},
195322  (2004).

\bibitem{Kikkawa1998}
J.~M. Kikkawa and D.~D. Awschalom, Phys. Rev. Lett. {\bf 80},  4313
(1998).

\bibitem{Ivchenko1973}
E.~L. Ivchenko, Sov. Phys. Solid State {\bf 15},  1048  (1973).

\bibitem{Chen2007}
Z. Chen, S.~G. Carter, R. Bratschitsch, P. Dawson, and S.~T.
Cundiff, Nature
  Phys. {\bf 3},  265  (2007).

\bibitem{StichB2007}
D. Stich, J. Zhou, T. Korn, R. Schulz, D. Schuh, W. Wegscheider,
M.~W. Wu, and
  C. Sch\"{u}ller, Phys. Rev. B {\bf 76},  205301  (2007).

\bibitem{Karimov2003}
O.~Z. Karimov, G.~H. John, R.~T. Harley, W.~H. Lau, M.~E. Flatt\'e,
M. Henini,
  and R. Airey, Phys. Rev. Lett. {\bf 91},  246601  (2003).

\bibitem{Sherman2005}
E.~Y. Sherman and J. Sinova, Phys. Rev. B {\bf 72},  075318  (2005).

\bibitem{Elliott1954}
R.~J. Elliott, Phys. Rev. {\bf 96},  266  (1954).

\bibitem{Tackeuchi1999}
A. Tackeuchi, T. Kuroda, S. Muto, Y. Nishikawa, and O. Wada, Jpn. J.
Appl. Phys
  {\bf 38},  4680  (1999).

\end{thebibliography}

\end{document}